# Age of Information Minimization in UAV-Assisted Covert Communication: Trajectory and Beamforming Design

Shima Salar Hosseini, Paeiz Azmi, *Senior Member, IEEE*, Ali Nazari

***Abstract*—Unmanned aerial vehicles (UAVs) have the potential for time-sensitive applications. Due to wireless channel variation, received data may have an expiration time, particularly in critical situations such as rescue operations, natural disasters, or the military. Age of Information (AoI) is a metric that measures the freshness of received packets to specify the validity period of information. In addition, it is necessary to guarantee the privacy of confidential information transmission through air-to-ground links against eavesdroppers. This paper investigates UAV-assisted covert communication to minimize AoI in the presence of an aerial eavesdropper for the first time. However, to ensure the eavesdropper's error detection rate, UAV-enabled beamforming employs the power-domain non-orthogonal multiple access (PD-NOMA) technique to cover the covert user by a public user. PD-NOMA technique significantly improves the user's AoI, too. The joint optimization problem contains non-convex constraints and coupled optimization variables, including UAV trajectory, beamforming design, and the user's AoI which is challenging to derive a direct solution. We have developed an efficient alternating optimization technique to address the formulated optimization problem. Numerical results demonstrate the impact of the main parameters on the performance of the proposed communication system.**

## I. Introduction

Unmanned Aerial Vehicles (UAVs) have the potential to significantly impact the deployment and operation of 5G and future 6G communication systems [1]. UAVs as aerial base stations (BS) can enhance network coverage and capacity where traditional ground-BS infrastructure is limited or unavailable. This can help improve network performance and connectivity in remote or disaster-stricken areas [2]. UAVs with edge capabilities can process and store data closer to the source, reducing latency and improving data transmission efficiency [3]. UAVs can improve security in sensitive, emergency, or high-risk environments to transfer sensitive data securely without the probability of interception or tampering [4]. Regardless of secure communication techniques i.e., encryption, and physical layer security, covert communication as an advanced level of security conceals the existence of confidential wireless transmission and avoids detecting transmission from eavesdroppers [5]. Based on the advantages of UAV mobility discussed above, integrating UAVs into a covert network can improve covertness by maneuvering towards legal receivers and maintaining distance from illegal receivers. In [6], in terms of maximizing the average covert transmission rate the UAV's trajectory, and transmit power are jointly optimized subject to transmission outage and covertness constraints with Willie's uncertain location. To maximize the average covert transmission rate under the constraints of UAVs' mobility, transmit powers, and warden's detection error probability, authors jointly optimized the UAVs' transmit powers and three-dimensional (3D) trajectories in [7]. In [8], the 3D trajectory and transmit power of UAV are jointly optimized to maximize the average covert transmission rate subject to the covertness constraint with active ground wardens. The scenario of eavesdropping on multiple wardens from the multiple UAV's links to ground users is considered in [9], where a UAV-mounted jammer generates artificial noise and assists covert communications. The problem is designed to max-min the average rate by jointly optimizing user association, bandwidth allocation, UAV transmit power control, and UAV 3D deployment, subject to the detection error probability of each warden's constraint. The air-to-ground (A2G) links on UAV-assisted covert communication systems cause the perfect detecting channels for an aerial warden, where most of the current articles focus on ground-based wardens, and only a few of them address the presence of an aerial warden. The authors in [10], proposed a UAV-relayed covert communication scheme with a ground transmitter and receiver with finite block length to maximize the effective transmission bits against a flying warden. The hovering location of the warden is obtained from the optimal detection thresholds for maximizing the covertness. Then, the block length and transmit power at the transmitter and the relay subject to the end-to-end error detection probability constraint are jointly optimized. Consequently, UAV-assisted covert communication suffers from high Willie's eavesdropping

S. Salar Hosseini, and P. Azmi are with the Department of Electrical and Computer Engineering, Tarbiat Modares University, Tehran, Iran e-mail: (shima.salarhosseini, and pazmi@modares.ac.ir)(Corresponding Author:Paeiz Azmi), Ali Nazari is with School of Electrical and Computer Engineering, College of Engineering, University of Tehran, Tehran, Iran (email: ali.nazary@ut.ac.ir).

due to free-space propagation signals. Beamforming is a promising approach for improving the covert rate with the capability of beamforming antenna design. In [11], the covert beamforming design for Internet-of-things (IoT) networks-assisted intelligent reflecting surfaces (IRS) is presented. To maximize the covert rate, Alice and IRS are jointly beamformers designed subject to the perfect covert transmission constraint, total transmit power constraint of Alice, and the quality-of-service (QoS) of the IRS. In [12], Alice communicates with Carol to cover the covert transmission to Bob, focusing on optimizing the beamformer design for enhanced covert transmission rates in a unicast beamforming network. The beamforming design problem to maximize the achievable covert rate under the perfect covert transmission constraint, the QoS of Carol, and the total transmit power constraints of Alice are jointly optimized. The authors in [13] demonstrate the equipping of Alice with an antenna array to perform 3D beamforming in the presence of the jammer with multiple antennas to improve the covert rate. In [14], the UAV satellite covert communication is considered to maximize the covert transmission rate by jointly optimizing the transmitter's 3D trajectory and 3D beamforming subject to the trajectory and covertness constraints. Consequently, one of the opportunistic techniques to guarantee covertness in the aerial system is proposing UAV-enabled beamforming to improve performance.

In addition, UAV-assisted covert communication can be utilized for emergency response and public safety applications, such as search and rescue missions, disaster assessment, and surveillance, and provide real-time situational awareness and support first responders in critical situations [15]. Enabling real-time coordination and decision-making is useful in military or government operations where confidentiality is paramount without requiring physical infrastructure. UAV's capabilities such as high mobility and fast deployment can be applied in various applications, including real-time monitoring, surveillance, agriculture, disaster response, and infrastructure inspection [16]. Therefore, UAVs can play an effective role in data freshness. Recently, a new metric named "age of information" is used for measuring data freshness, and refers to the time interval between signal generation and reaching the destination node [17]. Minimizing the age of information (AoI) in UAV-assisted covert networks ensures access to the most current data for effective decision-making. UAV missions can enhance the overall effectiveness and efficiency of freshness in different industries such as successful rescue operations, precision agriculture, natural disasters (earthquakes, hurricanes, or wildfires), and inspecting critical infrastructure (bridges, power lines, and pipelines) [18]. Investigation of information freshness in covert networks is a prominent area of interest in delay-sensitive secrecy applications. The authors in [19] jointly optimized the transmits probability and transmits the power of status information to maximize the covert energy-efficiency (EE) of the device-to-device (D2D) pair subject to the covertness and information freshness constraints. To minimize the average covert AoI under the covertness constraints, the authors in [20] determined the tradeoff between covertness and timeliness affected by the block length, transmit power, and prior transmission probability. The letter [21] addressed the requirement of information freshness, in the covertness maximization problem subject to the AoI constraint. In [22], the reliable covert communication problem in dynamic environments is demonstrated. To minimize AoI in the time-varying channels, the transmit power of Alice and the user's AoI are jointly optimized subject to the reliable covert constraint, the total transmit power constraint, the covertness constraint at Eve, and the QoS constraint of all users.

### *Motivation and Contribution*

UAVs with AoI freshness metrics have extensive applications that can support confidential scenarios, such as health assessments, timely medical interventions, identifying threats, and tracking movements. To the best of our knowledge, there has been no research on UAV-assisted covert communication to minimize AoI in the presence of an aerial eavesdropper. The proposed system model confronts two main challenges: i) the UAV's communication time related to the received packets' AoI of the users, and ii) due to the air-to-ground line-of-sight (LoS) channels, an aerial eavesdropper poses a serious threat to the security of covert communications. To address the mentioned challenges, this study investigates the joint design of UAV trajectory and beamforming to minimize the total AoI through the PD-NOMA transmission technique in the presence of an aerial eavesdropper. The main contributions of this paper are summarized as follows:

- We propose UAV-assisted covert communication using a beamforming technique in the PD-NOMA system for covert and public users against an aerial eavesdropper. In this context, we formulate UAV trajectory and beamforming design jointly to minimize the AoI, subject to the following constraints: power transmission budget, fairness in terms of guaranteeing the covert user, the covertness optimization, ensuring the user's packets reception before channel variations, quality of services, and UAV's maximum flying speed.
- An aerial eavesdropper makes a decision rule by jointly optimizing the distance to the UAV and the detection threshold. By employing uncertainty in UAV-enabled beamforming, we derive the eavesdropper's optimal detection error rate independent of the distance between the eavesdropper and the UAV. Therefore, the assumption of Willie's location uncertainty is unnecessary. Even in the worst-case situation, where an eavesdropper operates within the collision avoidance distance constraints of two UAVs, our analysis ensures that covertness.
- Additionally, to ensure covert communication of direct channels that are affected by the perfect eavesdropping, we have considered: i) UAV-assisted multiple antennas with beamforming design and uniform distribution power budget, and ii) a UAV applying the PD-NOMA technique to serve both public and covert users, thereby creating confusion for the eavesdropper through the superimposition of transmission power levels.

- Furthermore, by utilizing PD-NOMA, which employs SIC ordering, all users can receive their packets simultaneously without waiting in a queue. This approach effectively improves the AoI.
- The communication time in the proposed scenario depends on the packets' AoI of the users. Therefore, we discretize the communication time into time slots, with each duration guaranteeing the full reception of each packet before channel variation occurs. This leads to proposed effective constraints in trajectory design with a freshness approach.
- To tackle the proposed non-convex problems, we develop an alternating optimization approach. Hence, we decoupled our formulated optimization problem into three subproblems to obtain: 1) the AoI of the user which is in a linear programming standard form, 2) the UAV trajectory design which is approximated by the successive convex optimization technique, 3) the beamforming design which is approximated by semidefinite relaxation technique. Also, the non-convex constraints are approximated by the first-order Taylor expansion.
- Numerical results represent that, the proposed system achieved significant performance: 1) trajectory design and beamforming are both helpful in the achievable rate and AoI, 2) we always guarantee the achievable covert rate in the presence of an aerial eavesdropper, a) the UAV serving the public user by the PD-NOMA technique can effectively cover the covert user and degrade the error detection rate, b) serving the public user continuously during flying times while serves the covert user upon request (UAV monitors the covert user's requests and serves her/him as soon as possible at a desirable time to guarantee covert communication in a fresh manner.), 3) the adopted PD-NOMA technique outperforms the orthogonal multiple access scheme from the freshness of AoI, and achievable rate.

*Notations:* In this paper, scalars are denoted by italic letters, vectors and matrices are respectively represen ted by boldfaced lowercase and uppercase letters. $\mathbb{R}^{M\times 1}$, and $\mathbb{C}^{M\times 1}$ are denote the space of $M$-dimensional real-valued, and complex valued vector, respectively. $\mathbf{a}^T$, and $\mathbf{a}^H$ are transpose, and conjugate transpose of vector $\mathbf{a}$, respectively. Also, $|.|$ denotes the magnitude of a complex number, and $\|\ .\ \|$ denotes the Euclidean norm of vector. The expectation and the probability of $x$ are denoted by $\mathbb{E}\{x\}$, and $\Pr\{x\}$, respectively. $\mathcal{CN}(\mu, \sigma^2)$ denotes the complex Gaussian distribution with mean of $\mu$ and variance of $\sigma^2$.

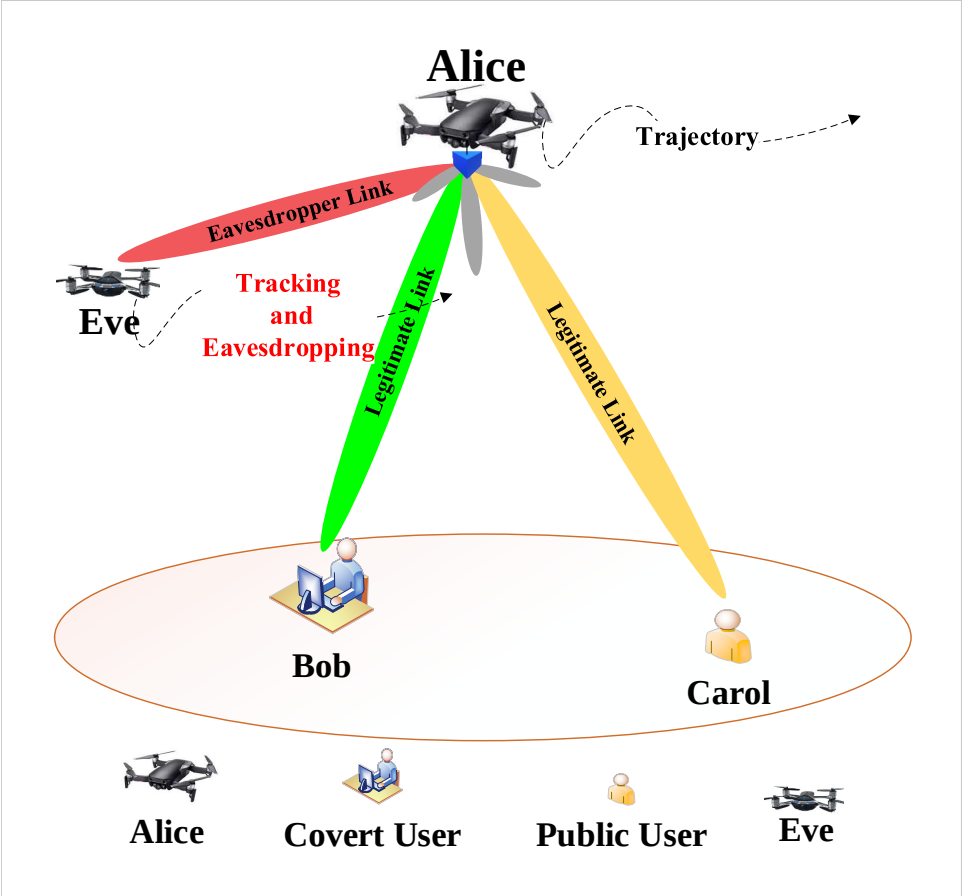

Fig. 1. The considered system model.

## II. System Model

### A. Considered Scenario and Assumptions

As illustrated in Fig. 1, we consider a UAV-assisted covert communication using a beamforming technique in the PD-NOMA system. The UAV, referred to as Alice, with beamforming serves the desired signal to the covert user (Bob) and public user (Carol) through the A2G channels while avoiding detection by the illegitimate user (Eve). Unlike most existing works [23]–[25], the probability of the presence of an aerial eavesdropper who can detect covert transmission through the A2G channels is considered. Alice is equipped with $M$ antennas, while legitimate and illegitimate receivers have a single antenna. Let $\mathcal{M} = \{1, \ldots, M\}$ denote the set of Alice's Antennas. It is a common assumption [26], for ease of exposition, we divide communication time $T$ into $N$ small equal time slots indexed by $n$ where $n \in \mathcal{N} = \{1, \ldots, N\}$. The location of Alice can be approximately unchanged in each time slot $n$ with length $\delta = T/N$ [27], even with maximum flying speed $V_{\max}$. Alice flies horizontally with two-dimensional (2D) Cartesian coordinates $\mathbf{q}[n] \triangleq [x[n], y[n]]^T \in \mathbb{R}^{2\times 1}$ at a constant altitude of $H$ above the ground in time slot $n$. Also, Alice's trajectory design is constrained by the maximum horizontal flying distance as follows:

$$\| \mathbf{q}[n+1] - \mathbf{q}[n] \| \leqslant V_{\max}\delta, \ \ n = 1, \ldots, N-1. \quad (1)$$

To verify the positive effect of PD-NOMA technique in covert communication, Alice superimposes the public signal $x_c^i[n]$ with covert signal $x_b^i[n]$ and then transmits from the $i$-th channel during the $n$-th time slot, as follows:

$$\mathbf{x}^i[n] = \mathbf{w}_c[n]x_c^i[n] + \mathbf{w}_b[n]x_b^i[n], \ \forall n, \quad (2)$$

here, $\mathbf{w}_k[n] \in \mathbb{C}^{M\times 1}$ represents the transmit beamforming vectors for the corresponding set $k \in \{b, c\}$, which refers to legitimate receivers. It is assumed that $\mathbb{E}\{|x_k^i[n]|^2\} = 1$, where $i = 1, \ldots, G$, and $G$ is the total number of channels used for transmitting $\mathbf{x}^i[n]$ to user $k$.

According to LoS communication links from Alice to the legitimate receivers, the channel gain from Alice to Bob,

and Carol during time slot $n$, which follows the free-space path loss model [28], can be expressed as:

$$\mathbf{h}_k[n] = \sqrt{\mu_0 d_k^{-2}[n]}\mathbf{a}(\mathbf{q}[n], \mathbf{u}_k), \tag{3}$$

where $d_k[n] = \sqrt{\| \mathbf{q}[n] - \mathbf{u}_k \|^2 + H^2}$ refers to the distance between Alice and the legitimate user in the $n$th time slot, and $\mathbf{u}_k = [x_k, y_k] \in \mathbb{R}^{2\times 1}$ is the 2D Cartesian coordinates of legitimate ground users. The channel power at the reference distance $1m$, is denoted by $\mu_0$. Additionally, $\mathbf{a}(\mathbf{q}[n], \mathbf{u}_k)$ represents the transmit array response vector of Alice toward the legitimate user $k$, expressed as:

$$\begin{aligned}\mathbf{a}(\mathbf{q}[n], \mathbf{u}_k) = [1, & e^{-j2\pi\frac{d}{\lambda}\sin(\theta(\mathbf{q}[n],\mathbf{u}_k))}, \\ & \dots, e^{-j2\pi\frac{d}{\lambda}(M-1)\sin(\theta(\mathbf{q}[n],\mathbf{u}_k))}]^T, \end{aligned}\tag{4}$$

where $d$, and $\lambda$ are the space between two adjacent antennas and the carrier wavelength, respectively. Furthermore, $\sin(\theta(\mathbf{q}[n], \mathbf{u}_k)) = \frac{H}{d_k[n]}$, that $\theta$ is the angle of departure (AoD) from Alice corresponding to legitimate user $k$.

In the time slot $n$ where Alice serves Bob and Carol, the achievable data rates can be respectively denoted as follows:

$$R_b[n] = \log_2\left(1 + \frac{\left|\mathbf{h}_b^H[n]\mathbf{w}_b[n]\right|^2}{\sigma_b^2}\right), \tag{5}$$

$$R_c[n] = \log_2\left(1 + \frac{\left|\mathbf{h}_c^H[n]\mathbf{w}_c[n]\right|^2}{\left|\mathbf{h}_c^H[n]\mathbf{w}_b[n]\right|^2 + \sigma_c^2}\right), \tag{6}$$

where $\sigma_k^2$ represents the power of additive white Gaussian noise (AWGN) at legitimate user $k$, where $k \in \{b, c\}$.

### B. Binary Hypothesis Testing at Eve

Eve attempts to detect covert signal transmissions from A2G links using a radiometer. Also, Eve's horizontal coordinate is assumed $\mathbf{l}[n] \triangleq [x_e[n], y_e[n]]^T \in \mathbb{R}^{2\times 1}$ at a constant altitude $h$ above the ground in time slot $n$. Regarding the aerial eavesdropping link between Alice and Eve, the channel gain during time slot $n$ follows a large-scale LoS path loss [10], and is expressed as:

$$\mathbf{h}_e[n] = \sqrt{\mu_0 d_e^{-2}[n]}\mathbf{a}(\mathbf{q}[n], \mathbf{l}[n]), \tag{7}$$

where $d_e[n] = \sqrt{\| \mathbf{q}[n] - \mathbf{l}[n] \|^2 + (H-h)^2}$ refers to the distance between Alice and Eve in the $n$-th time slot. Similarly to formulation (4), the eavesdropping array response vector of Alice toward Eve is defined as:

$$\begin{aligned}\mathbf{a}(\mathbf{q}[n], \mathbf{l}[n]) = [1, & e^{-j2\pi\frac{d}{\lambda}\sin(\phi(\mathbf{q}[n],\mathbf{l}[n]))}, \\ & \dots, e^{-j2\pi\frac{d}{\lambda}(M-1)\sin(\phi(\mathbf{q}[n],\mathbf{l}[n]))}]^T, \end{aligned}\tag{8}$$

where $\sin(\phi(\mathbf{q}[n], \mathbf{l}[n])) = \frac{H-h}{d_e[n]}$ and $\phi$ is the AoD of the eavesdropping link from Alice to Eve.

*1) Detection Threshold Analysis:* The received signal at Eve under two hypothesis tests: null $\mathcal{H}_0$, and the alternative $\mathcal{H}_1$, is demonstrated as follows:

$$y_e^i[n] = \begin{cases}\mathbf{h}_e^H[n]\mathbf{w}_c[n]x_c^i[n] + n_e^i, & \mathcal{H}_0, \\ \mathbf{h}_e^H[n](\mathbf{w}_c[n]x_c^i[n] + \mathbf{w}_b[n]x_b^i[n]) + n_e^i, & \mathcal{H}_1,\end{cases} \tag{9}$$

where $n_e^i \sim \mathcal{CN}(0, \sigma_e^2)$ is the AWGN at Eve from $i$-th channel with zero mean and variance $\sigma_e^2$.

Based on (9), Eve decides Alice's covert transmission. Hence, the optimal decision rule to minimize the error detection rate at Eve can be expressed as follows:

$$T_e[n] = \frac{1}{G}\sum_{i=1}^{G}\left|y_e^i[n]\right|^2 \underset{\mathcal{D}_0}{\overset{\mathcal{D}_1}{\gtrless}} \tau[n], \tag{10}$$

where $T_e[n]$ is the average power of received signal from Alice to Eve, $\tau[n]$ is the detection threshold at time slot $n$, and $\mathcal{D}_0$ and $\mathcal{D}_1$ are the decision parameters in favor of $\mathcal{H}_0$ and $\mathcal{H}_1$, respectively. It is common in UAV network literature to assume an infinite number of channel links in each time slot [6]. Similarly, in this scenario, we consider that Eve can receive signals from an infinite number of channel links, i.e., $i \to \infty$. By noting that $x_c^i[n]$, $x_b^i[n]$, and $n_e^i[n]$ are independent, and based on (9), and (10), $T_e[n]$ is rewritten as follows:

$$T_e[n] = \begin{cases}\left|\mathbf{h}_e^H[n]\mathbf{w}_c[n]\right|^2 + \sigma_e^2, & \mathcal{H}_0, \\ \left|\mathbf{h}_e^H[n]\mathbf{w}_c[n]\right|^2 + \left|\mathbf{h}_e^H[n]\mathbf{w}_b[n]\right|^2 + \sigma_e^2, & \mathcal{H}_1.\end{cases} \tag{11}$$

The performance of the hypothesis testing at Eve to minimize the detection error rate $\xi[n]$ at time slot $n$, achieved from two probabilities of false alarm $\mathbb{P}_{\text{FA}}[n] = \Pr\{\mathcal{D}_1|\mathcal{H}_0\}$, and miss detection $\mathbb{P}_{\text{MD}}[n] = \Pr\{\mathcal{D}_0|\mathcal{H}_1\}$, as follows:

$$\xi[n] = \mathbb{P}_{\text{FA}}[n] + \mathbb{P}_{\text{MD}}[n], \ \forall n. \tag{12}$$

In covert communications with an aerial eavesdropper, Eve aims to minimize the detection error rate under the optimal detection threshold value $\tau[n]$ and optimal hovering location $\mathbf{l}[n]$. Therefore, we should first derive the minimum optimal detection error rate $\xi^*[n]$ from Eve's perspective. Afterward, Alice jointly designs the trajectory and beamforming vectors to obtain the optimal covert rate.

*2) The Performance of Error Detection Probability:* Based on the average received power at Eve presented in (11), the false alarm and miss detection probabilities of the proposed UAV-assisted covert communication using a beamforming technique in the PD-NOMA system are derived as:

$$\mathbb{P}_{\text{FA}}[n] = \Pr\{\left|\mathbf{h}_e^H[n]\mathbf{w}_c[n]\right|^2 + \sigma_e^2 > \tau[n]\}, \tag{13}$$

$$\mathbb{P}_{\text{MD}}[n] = \Pr\{\left|\mathbf{h}_e^H[n]\mathbf{w}_c[n]\right|^2 + \left|\mathbf{h}_e^H[n]\mathbf{w}_b[n]\right|^2 + \sigma_e^2 < \tau[n]\}. \tag{14}$$

Motivated to guarantee covert communication, we employed a UAV-enabled beamforming design to increase Eve's uncertainties. Since the beamformer vector's $\mathbf{w}_k[n]$ are designed based on the Channel State Information (CSI) of legitimate links, deriving the detection error rate

$$\mathbb{P}_{\mathrm{MD}}[n] = \Pr\{|\alpha_c[n]|^2 + |\alpha_b[n]|^2 + \sigma_e^2 < \tau[n]\} = \begin{cases} \frac{\varpi_b[n]\exp\left(\frac{\sigma_e^2-\tau[n]}{\varpi_b[n]}\right) - \varpi_c[n]\exp\left(\frac{\sigma_e^2-\tau[n]}{\varpi_c[n]}\right)}{\varpi_c[n]-\varpi_b[n]} + 1, & \tau[n] > \sigma_e^2, \\ 0, & \tau[n] < \sigma_e^2. \end{cases} \quad (18)$$

$$\xi[n] = \begin{cases} \frac{\varpi_b[n]}{\varpi_c[n]-\varpi_b[n]}\left[\exp\left(\frac{\sigma_e^2-\tau[n]}{\varpi_b[n]}\right) - \exp\left(\frac{\sigma_e^2-\tau[n]}{\varpi_c[n]}\right)\right] + 1, & \tau[n] > \sigma_e^2, \\ 1, & \tau[n] < \sigma_e^2. \end{cases} \quad (19)$$

is challenging for Eve. Eve assumes that Alice designs the beamformer for each antenna using $w_{k,m}[n] = w^r_{k,m}[n] + jw^i_{k,m}[n]$, where $w^r_{k,m}[n]$ and $w^i_{k,m}[n]$ are independent and identically distributed (i.i.d.) with normal distributions, specifically $w^r_{k,m}[n] \sim \mathcal{N}(0, \frac{\sigma^2_{k,m}}{2})$ and $w^i_{k,m}[n] \sim \mathcal{N}(0, \frac{\sigma^2_{k,m}}{2})$. In addition, the beamforming vectors $\mathbf{w}_k[n]$ are independently and jointly with complex Gaussian distributions $\mathbf{w}_k[n] \sim \mathcal{CN}(\mathbf{0}, \mathbf{\Sigma})$, where $\mathbf{0}$ represents the zero-mean vector, and $\mathbf{\Sigma}$ denotes the covariance matrix as follows:

$$\mathbf{\Sigma} = \begin{bmatrix} \sigma^2_{k,1}[n] & 0 & \dots & 0 \\ 0 & \sigma^2_{k,2}[n] & \dots & 0 \\ \vdots & \vdots & \ddots & \vdots \\ 0 & 0 & \dots & \sigma^2_{k,M}[n] \end{bmatrix}, \quad (15)$$

here, $\sigma^2_{k,m}[n] = E\{|\mathrm{w}_{k,m}[n]|^2\}$ is the variance of $\mathrm{w}_{k,m}[n]$, $k \in \{b, c\}$, $m \in \mathcal{M}$. Eve characterizes the distribution function of the false alarm and miss detection probabilities by denoting $\alpha_k[n]$ as follows:

$$\alpha_k[n] = \mathbf{h}^H_e[n]\mathbf{w}_k[n] = \sum_{m=1}^{M} h^*_{e,m}[n]\mathrm{w}_{k,m}[n], k \in \{b, c\}, \quad (16)$$

where the distribution of $\alpha_k[n]$ is analyze in the following lemma.

*Theorem:* The sum of independent normally distributed random variables follows a normal distribution [29].

*Lemma 1:* Following the theorem, since the $m$-th element of beamforming vector to $k$-th user follows a complex normal distribution $\mathrm{w}_{k,m}[n] \sim \mathcal{CN}(0, \sigma^2_{k,m}[n])$, the distribution of $\alpha_k[n]$ can be determined as $\alpha_k[n] \sim \mathcal{CN}(0, \varpi_k[n])$, where $\varpi_k[n] = \sum_{m=1}^{M} |h_{e,m}[n]|^2\sigma^2_{k,m}[n]$. In addition, $|\alpha_k[n]|^2$ follow an exponential distribution, i.e., $|\alpha_k[n]|^2 \sim \exp\left(\frac{1}{\varpi_k[n]}\right)$.

*Proof* : The detailed proof is provided in Appendix A.■

Consequently, based on (13), (16) the false alarm probability $\mathbb{P}_{\mathrm{FA}}[n]$ of the proposed scheme at Eve can be derived as follows:

$$\begin{aligned}\mathbb{P}_{\mathrm{FA}}[n] &= \Pr\{|\alpha_c[n]|^2 + \sigma_e^2 > \tau[n]\}, \\ &= \begin{cases} \exp\left(\frac{\sigma_e^2-\tau[n]}{\varpi_c[n]}\right), & \tau[n] > \sigma_e^2, \\ 1, & \tau[n] < \sigma_e^2. \end{cases}\end{aligned} \quad (17)$$

The miss detection probability $\mathbb{P}_{\mathrm{MD}}[n]$ at Eve is obtained from (14) and (16), in (18). Therefore, the error detection rate $\xi[n]$ is obtained by substituting equations (17) and (18) into equation (12) in equation (19).

### C. Age of Information

In the proposed system, we consider Alice flies to serve Bob and Carol with the PD-NOMA technique. However, Alice faces communication time constraints due to limited onboard UAV energy. On the other hand, in time-sensitive applications, the timeliness of received data is important. Therefore, we leverage the information freshness metrics in the context of covert communication in time-varying channels. The age of information refers to the elapsed time from the received packet by the legitimate users that has been generated at Alice, which is defined as follows:

$$\Delta_k(t) = t - \iota_k(t), \; k \in \{b, c\}, \quad (20)$$

where $\iota_k(t)$ refers to the time the most recently received packet at the $k$-th legitimate user, was generated at Alice. For ease of exposition, we assume $\Delta_k(0) = 0$. To enhance the AoI, we utilize the first come first served (FCFS) method to update the age status [30]. In FCFS systems, a new packet's transmission is available exactly at the transmitter when the packet's update in service finishes at the destination. In this manner, the waiting time for updating the packet is almost near zero which will obtain the smallest age of freshness that aligns with the aim of the proposed system model to minimize the AoI.

## III. Problem Formulation and Solution Methodology

### A. Problem Formulation

In the context of UAV-assisted covert communication using beamforming in the PD-NOMA scheme, our objective is to minimize the total AoI among all legitimate users. This is achieved by jointly optimizing the trajectory of Alice (denoted as $\mathbf{Q} = \{\mathbf{q}[n], \forall n\}$), transmit beamformers (denoted as $\mathbf{W} = \{\mathbf{w}_k[n], \forall k, n\}$), and the freshness of information (denoted as $\mathbf{\Delta} = \{\Delta_k[n], \forall k, n\}$) over all time slots. The jointly optimization problem is formulated as:

$$\min_{\mathbf{Q},\mathbf{W},\boldsymbol{\Delta}} \sum_{n=1}^{N} \sum_{k\in\{b,c\}} \Delta_k[n] \tag{21a}$$

$$\text{s.t.} \sum_{k\in\{b,c\}} \parallel \mathbf{w}_k[n] \parallel^2 \leqslant \mathit{\Gamma},\ \forall n, \tag{21b}$$

$$|\mathbf{h}_k^H[n]\mathbf{w}_c[n]|^2 > |\mathbf{h}_k^H[n]\mathbf{w}_b[n]|^2,\ k\in\{b,c\},\forall n, \tag{21c}$$

$$\min_{\tau[n],\mathbf{l}[n]} \xi[n] \geqslant 1-\epsilon,\ \forall n, \tag{21d}$$

$$\max_k \Delta_k[n] \leqslant \delta,\ \forall n, \tag{21e}$$

$$\Delta_c[n] \times R_c[n] \geqslant \frac{S_c[n]}{B},\ \forall n, \tag{21f}$$

$$\sum_{n=1}^{N} (\Delta_b[n] \times R_b[n]) \geqslant \frac{S_b}{B}, \tag{21g}$$

$$\parallel \mathbf{q}[n+1]-\mathbf{q}[n] \parallel \leqslant \max_k \Delta_k[n] \times V_{\max},\quad n=1,\ldots,N-1, \tag{21h}$$

where the beamformer vectors $\mathbf{w}_k[n]$ satisfy (21b), that $\mathit{\Gamma}$ represents the transmit power of Alice with a uniform distribution, subject to an upper bound of $P_{\max}$. In (21c), we achieve fairness in the NOMA scheme to jointly improve covertness and AoI. This is realized by allocating more power to Carol and enabling the successful implementation of SIC at Bob [31]. The constraint (21d) ensures that the detection error rate minimization problem is not less than a specific value. Constraint (21e) ensures that the maximum age of packet freshness must be shorter than the duration of each time slot, actually before channel variation. The QoS constraints (21f) and (21g) ensure successful packet transmissions from Alice to Bob and Carol with, minimum required sizes $S_b$ and $S_c[n]$, respectively. $B$ denotes the communication bandwidth link. The constraint (21h) guarantees that the total packet is transmitted within the maximum horizontal flight distance of Alice across all time slots.

*Lemma 2:* To enhance the tractability of the optimization problem, we initially address the optimization constraint (21d) to determine Eve's optimal detection threshold $\tau^*[n]$ and optimal location $\mathbf{l}[n]$, resulting achieving the minimum detection error rate $\xi^*[n]$.

*Proof* : The detailed proof is reported in Appendix B.

In the subsequent sections, leveraging Lemma 2, we employ $\xi^*[n] \geqslant 1-\epsilon$ instead of (21d) as the covertness constraint in our joint optimization problem.

The joint optimization problem (21) is difficult to solve because the UAV trajectory variables $\mathbf{Q}$, beamforming design variables $\mathbf{W}$, and AoI variables $\boldsymbol{\Delta}$ are strongly coupled in the constraints. Furthermore, the covertness constraint (21d), QoS constraints (21f), and (21g) are non-convex, and complicating the solution process. Therefore, to address the non-convex formulated problem, we decompose the joint optimization problem (21) into three subproblems: AoI freshness optimization, Alice trajectory design, and beamforming design optimization. Subsequently, we developed an efficient alternative optimization algorithm by adopting successive convex (SC) optimization techniques.

### *B. Solution Methodology*

*1) AoI Optimization:* To obtain the optimal AoI freshness from optimization problem (21) for a given UAV trajectory design and transmit beamformers $\{\mathbf{Q},\mathbf{W}\}$, we solve the following optimization problem:

$$\min_{\boldsymbol{\Delta}} \sum_{n=1}^{N} \sum_{k\in\{b,c\}} \Delta_k[n] \tag{22a}$$

$$\text{s.t.} \max_k \Delta_k[n] \leqslant \delta,\ \forall n, \tag{22b}$$

$$\Delta_c[n] \times R_c[n] \geqslant \frac{S_c[n]}{B},\ \forall n, \tag{22c}$$

$$\sum_{n=1}^{N} (\Delta_b[n] \times R_b[n]) \geqslant \frac{S_b}{B}, \tag{22d}$$

$$\parallel \mathbf{q}[n+1]-\mathbf{q}[n] \parallel \leqslant \max_k \Delta_k[n] \times V_{\max},\quad n=1,\ldots,N-1, \tag{22e}$$

since problem (22) and its constraints are in standard linear programming (LP) form, it can be efficiently solved using existing optimization tools like CVX [32].

*2) Trajectory Design Optimization:* The optimal UAV trajectory design for a specified AoI freshness, and transmit beamformers $\{\boldsymbol{\Delta},\mathbf{W}\}$ can be obtained by solving the following optimization problem:

$$\min_{\mathbf{Q}} \sum_{n=1}^{N} \sum_{k\in\{b,c\}} \Delta_k[n] \tag{23a}$$

$$\text{s.t.} \Delta_c[n] \times R_c[n] \geqslant \frac{S_c[n]}{B},\ \forall n, \tag{23b}$$

$$\sum_{n=1}^{N} (\Delta_b[n] \times R_b[n]) \geqslant \frac{S_b}{B}, \tag{23c}$$

$$\parallel \mathbf{q}[n+1]-\mathbf{q}[n] \parallel \leqslant \max_k \Delta_k[n] \times V_{\max},\quad n=1,\ldots,N-1, \tag{23d}$$

where (23) is a non-convex optimization problem due to the non-convex constraints (23b) and (23c). Accordingly, we adopt a successive convex approximation (SCA) approach to iteratively determine the optimal trajectory design of Alice. On the other hand, to enhance the signal energy and improve the achievable rate, the phase angles $\theta$ of all beamformer vectors can be jointly adjusted to achieve phase alignment of signals from different transmission paths at legitimate users. Therefore, we express the $\mathbf{h}_b^H[n]\mathbf{w}_b[n]$ at formulation (24). Hence, the following upper bound is provided for the term $\mathbf{h}_b^H[n]\mathbf{w}_b[n]$:

$$\mathbf{h}_b^H[n]\mathbf{w}_b[n] \leqslant \frac{\sqrt{\mu_0}}{d_b[n]} \sum_{m=1}^{M} |\mathrm{w}_b^m[n]|. \tag{25}$$

$$\mathbf{h}_b^H[n]\mathbf{w}_b[n] = \frac{\sqrt{\mu_0}\sum_{m=1}^{M}|\mathrm{w}_b^m[n]|e^{j\left(\frac{2(m-1)\pi d}{\lambda}\sin(\theta(\mathbf{q}[n],\mathbf{u}_k))+\measuredangle \mathrm{w}_b^m[n]\right)}}{d_b[n]}. \tag{24}$$

Consequently, the term $|\mathbf{h}_b^H[n]\mathbf{w}_b[n]|^2$, has the following upper bound:

$$|\mathbf{h}_b^H[n]\mathbf{w}_b[n]|^2 \leqslant \frac{\mu_0 z_b[n]}{d_b^2[n]}, \tag{26}$$

where $z_k[n] = (\sum_{m=1}^{M}|\mathrm{w}_k^m[n]|)^2,\ k\in\{b,c\}$. Similarly, for $|\mathbf{h}_c^H[n]\mathbf{w}_k[n]|^2$, we express the upper bound as follow:

$$|\mathbf{h}_c^H[n]\mathbf{w}_k[n]|^2 \leqslant \frac{\mu_0 z_k[n]}{d_c^2[n]},\ k\in\{b,c\}. \tag{27}$$

Therefore, the achievable data rate at legitimate users are:

$$\hat{R}_b[n] = \log_2\left(1+\frac{\eta z_b[n]}{j_b[n]+H^2}\right), \tag{28}$$

$$\hat{R}_c[n] = \log_2\left(1+\frac{\eta z_c[n]}{\eta z_b[n]+j_c[n]+H^2}\right), \tag{29}$$

where $\eta = \mu_0/\sigma^2$, and inequality $j_k[n] \leqslant\parallel \mathbf{q}[n]-\mathbf{u}_k \parallel^2$ , $k\in\{b,c\}$ are facilitate the derivation of upper bound for the concave data rate function through its first-order Taylor expansion at any point:

$$\begin{aligned}\check{R}_b[n] =& \log_2\left(\eta z_b[n]+j_b[n]+H^2\right)-\log_2\left(j_b^{\varsigma}[n]+H^2\right)\\ &-\frac{\log_2(e)}{j_b^{\varsigma}[n]+H^2}\left(j_b[n]-j_b^{\varsigma}[n]\right),\end{aligned} \tag{30}$$

$$\begin{aligned}\check{R}_c[n] =& \log_2\left(\eta(z_b[n]+z_c[n])+j_c[n]+H^2\right)\\ &-\log_2\left(\eta z_b[n]+j_c^{\varsigma}[n]+H^2\right)\\ &-\frac{\log_2(e)}{\eta z_b[n]+j_c^{\varsigma}[n]+H^2}\left(j_c[n]-j_c^{\varsigma}[n]\right).\end{aligned} \tag{31}$$

Since, $j_k[n]$ is a slack variable, we have the following inequality by applying the first-order Taylor expansion at the given point $q^{\varsigma}[n]$ for $\varsigma$-th iteration:

$$\begin{aligned}j_k[n] \leqslant\parallel q^{\varsigma}[n]-\mathbf{u}_k \parallel^2 &+2(q^{\varsigma}[n]-\mathbf{u}_k)^T(\mathbf{q}[n]-q^{\varsigma}[n]),\\ &k\in\{b,c\},\ \forall n.\end{aligned} \tag{32}$$

Consequently, the non-convex optimization problem (23) is replaced with the following convex optimization problem:

$$\min_{\mathbf{Q}}\sum_{n=1}^{N}\sum_{k\in\{b,c\}}\Delta_k[n] \tag{33a}$$

$$\text{s.t.}\Delta_c[n]\times\check{R}_c[n]\geqslant\frac{S_c[n]}{B},\ \forall n, \tag{33b}$$

$$\sum_{n=1}^{N}\left(\Delta_b[n]\times\check{R}_b[n]\right)\geqslant\frac{S_b}{B}, \tag{33c}$$

$$(23d),\ (32). \tag{33d}$$

Problem (33) is a convex optimization problem that can be effectively solved using standard solvers like CVX [32].

*3) Beamforming Optimization:* For any given AoI data freshness and as well as UAV trajectory design $\{\mathbf{\Delta},\mathbf{Q}\}$, the transmit beamformers of problem (21) can be optimized by solving the following problem:

$$\min_{\mathbf{W}}\sum_{n=1}^{N}\sum_{k\in\{b,c\}}\Delta_k[n] \tag{34a}$$

$$\text{s.t.}\sum_{k\in\{b,c\}}\parallel \mathbf{w}_k[n]\parallel^2\leqslant\mathit{\Gamma},\ \forall n, \tag{34b}$$

$$|\mathbf{h}_k^H[n]\mathbf{w}_c[n]|^2 > |\mathbf{h}_k^H[n]\mathbf{w}_b[n]|^2,\ k\in\{b,c\},\forall n, \tag{34c}$$

$$\xi^*[n]\geqslant 1-\epsilon,\ \forall n, \tag{34d}$$

$$\Delta_c[n]\times R_c[n]\geqslant\frac{S_c[n]}{B},\ \forall n, \tag{34e}$$

$$\sum_{n=1}^{N}\left(\Delta_b[n]\times R_b[n]\right)\geqslant\frac{S_b}{B}. \tag{34f}$$

While the objective function (34) and the constraints (34b) and (34c) are convex, it is challenging to achieve the optimal transmit beamformers due to the coupling in non-convex constraints (34d), (34e), and (34f). We apply the semidefinite relaxation (SDR) and SCA alternating optimization techniques to solve the problem, respectively. Let we define $\mathbf{H}_k[n]=\mathbf{h}_k[n]\mathbf{h}_k^H[n]$, and $\mathbf{W}_k[n]=\mathbf{w}_k[n]\mathbf{w}_k^H[n]$, where $\mathbf{W}_k[n]\geqslant 0$ and $\text{rank}(\mathbf{W}_k[n])=1,\ k\in\{b,c\}$. In addition, $p_k[n]=\text{Tr}(\mathbf{W}_k[n])$, and $|\mathbf{h}_k^H[n]\mathbf{w}_k[n]|^2=\text{Tr}\left(\mathbf{H}_k[n]\mathbf{W}_k[n]\right)$. Accordingly, problem (34) is reformulated as:

$$\min_{\mathbf{W}}\sum_{n=1}^{N}\sum_{k\in\{b,c\}}\Delta_k[n] \tag{35a}$$

$$\text{s.t.}\sum_{k\in\{b,c\}}p_k[n]\leqslant\mathit{\Gamma},\ \forall n, \tag{35b}$$

$$\text{Tr}\left(\mathbf{H}_k[n]\mathbf{W}_c[n]\right)\geqslant\text{Tr}\left(\mathbf{H}_k[n]\mathbf{W}_b[n]\right),\ k\in\{b,c\},\forall n, \tag{35c}$$

$$\Upsilon\left(p_b[n],p_c[n]\right)\leqslant\epsilon,\forall n, \tag{35d}$$

$$\frac{1}{f_k[n]}\leqslant\text{Tr}\left(\mathbf{H}_k[n]\mathbf{W}_k[n]\right),\quad k\in\{b,c\},\ \forall n, \tag{35e}$$

$$g_k[n]\geqslant\sum_{\Omega(i)>\Omega(k)}\text{Tr}(\mathbf{H}_k[n]\mathbf{W}_i[n])+\sigma_k^2,\ k\in\{b,c\},\ \forall n, \tag{35f}$$

$$\Delta_c[n]\times\log_2\left(1+\frac{1}{f_c[n]g_c[n]}\right)\geqslant\frac{S_c[n]}{B},\ \forall n, \tag{35g}$$

$$\sum_{n=1}^{N}\left(\Delta_b[n]\times\log_2\left(1+\frac{1}{f_b[n]g_b[n]}\right)\right)\geqslant\frac{S_b}{B}, \tag{35h}$$

$$\mathbf{W}_k[n]\succeq 0,\ \forall n, \tag{35i}$$

$$\text{rank}\left(\mathbf{W}_k[n]\right)=1,\forall n, \tag{35j}$$

where $\Upsilon(p_b[n],p_c[n]) = \frac{p_b[n]}{p_c[n]-p_b[n]}\times\left[\left(\frac{p_c[n]}{p_b[n]}\right)^{\frac{p_b[n]}{p_b[n]-p_c[n]}}-\left(\frac{p_c[n]}{p_b[n]}\right)^{\frac{p_c[n]}{p_b[n]-p_c[n]}}\right]$, and $\Omega(k)$ specifies the decoding order

$$\Upsilon(p_b^\iota[n], p_c^\iota[n]) + \sum_{k\in\{b,c\}} (\frac{\partial \Upsilon(p_b^\iota[n], p_c^\iota[n])}{\partial p_k^\iota[n]}) \times (p_k[n] - p_k^\iota[n]) \leqslant \epsilon \tag{36}$$

where

$$\frac{\partial \Upsilon(p_b^\iota[n], p_c^\iota[n])}{\partial p_b^\iota[n]} = \frac{p_c^\iota[n]}{(p_c^\iota[n]-p_b^\iota[n])^2} \times [(\frac{p_c^\iota[n]}{p_b^\iota[n]})^{\frac{p_b^\iota[n]}{p_b^\iota[n]-p_c^\iota[n]}} - (1+\ln(\frac{p_c^\iota[n]}{p_b[n]^\iota}))(\frac{p_c^\iota[n]}{p_b^\iota[n]})^{\frac{p_c^\iota[n]}{p_b^\iota[n]-p_c^\iota[n]}}], \tag{37}$$

$$\frac{\partial \Upsilon(p_b^\iota[n], p_c^\iota[n])}{\partial p_c^\iota[n]} = \frac{-p_b^\iota[n]}{(p_c^\iota[n]-p_b^\iota[n])^2} \times [(\frac{p_c^\iota[n]}{p_b^\iota[n]})^{\frac{p_b^\iota[n]}{p_b^\iota[n]-p_c^\iota[n]}} - (1+\ln(\frac{p_c^\iota[n]}{p_b^\iota[n]}))(\frac{p_c^\iota[n]}{p_b^\iota[n]})^{\frac{p_c^\iota[n]}{p_b^\iota[n]-p_c^\iota[n]}}]. \tag{38}$$

for user $k$, where $\Omega(i) > \Omega(k)$ indicates that user $k$ has a smaller order and therefore detects its signal earlier. Based on the non-convex constraints (35d), (35g), (35h), and (35j), the problem (35) is a non-convex. Hence, by applying the first-order Taylor expansion to $\Upsilon(p_b[n], p_c[n])$ at the given points $p_b^\iota[n]$, and $p_c^\iota[n]$ in the $\iota$-th iteration, the optimal solution is obtained at (36)-(38). However, the constraints (35g) and (35h) still lead to non-convexity of problem formulation (35). With respect to $x$ and $y$, for $x > 0$, and $y > 0$, $f(x,y) = \log\left(1+\frac{1}{xy}\right)$ is a joint convex function [33]. Therefore, term $\log_2\left(1+\frac{1}{f_k[n]g_k[n]}\right)$ is joint convex function over $f_k[n]$, and $g_k[n]$. Hence, the first-order Taylor expansion can be used to linearly approximate an upper bound at given local points $f_k^\iota[n]$, and $g_k^\iota[n]$ to generate a tighter convex substitute:

$$\begin{aligned} &\log_2(1+\frac{1}{f_k[n]g_k[n]}) \geqslant \log_2(1+\frac{1}{f_k^\iota[n]g_k^\iota[n]}) \\ &- \frac{\log_2(e)(f_k[n]-f_k^\iota[n])}{f_k^\iota[n](1+f_k^\iota[n]g_k^\iota[n])} - \frac{\log_2(e)(g_k[n]-g_k^\iota[n])}{g_k^\iota[n](1+f_k^\iota[n]g_k^\iota[n])}, \\ &= \tilde{R}_k[n],\ k\in\{b,c\},\ \forall n. \end{aligned} \tag{39}$$

Regarding to the non-convex constraint (35j), an optimal solution is always obtained by satisfing the rank-one constraint [33], [34]. Consequently, the optimization problem (35) can be reformulated as follows:

$$\min_{\mathbf{W}} \sum_{n=1}^{N} \sum_{k\in\{b,c\}} \Delta_k[n] \tag{40a}$$

$$\text{s.t.} \Delta_c[n] \times \tilde{R}_c[n] \geqslant \frac{S_c[n]}{B},\ \forall n, \tag{40b}$$

$$\sum_{n=1}^{N} \left(\Delta_b[n] \times \tilde{R}_b[n]\right) \geqslant \frac{S_b}{B}, \tag{40c}$$

$$\text{(35b), (35c), (36), (35e), (35f), (35i).} \tag{40d}$$

Consequently, the formulation presented in (40) qualifies as a convex semidefinite program (SDP) that can be effectively tackled with standard convex optimization tools, such as CVX [32].

## IV. Numerical Results

The numerical results demonstrate the potential performance of UAV-assisted covert communication using a beamforming technique within a PD-NOMA framework, even in the presence of an aerial eavesdropper. To ensure covert communication, Alice continuously serves Carol while providing service to Bob upon his request during desirable time slots. Bob and Carol are uniformly and randomly distributed in the 2D area of $1 \times 1$ km$^2$. Alice is equipped with $M = 10$ antennas and flies at a fixed altitude of $H = 100$ m with the maximum speed of $V_{\max} = 30$ m/s. The minimum required packet sizes for Bob and Carol are set to $S_b = 45$ Mbit, and $S_c[n] = 5$ Mbit, respectively. The channel power gain is characterized by $\mu_0 = -30$ dB. The noise power for legal and illegal receivers is given as $\sigma_k^2 = \sigma_w^2 = -100$ dB, $k \in \{b,c\}$. The antenna spacing is set as half of a wavelength. Other parameters include $B = 1$ MHz.

### A. Impact of the number of Alice Antennas

The achievable rates versus the different numbers of antennas $M$ are represented in Fig. 2. As shown for a given $M$, Carol's achievable rate is more than Bob which ensures covert communication due to the following reasons: 1) Alice serves Carol in all time slots while she transmits to Bob upon his requests in desirable time slots, 2) Alice employs a superposition coding strategy for covert and public signals using the PD-NOMA technique with fairness constraint. Furthermore, increasing the antennas leads to improving the corresponding achievable data rate. On the other hand, the achievable rate for Carol and Bob increases as the transmit power of Alice $\Gamma$ increases.

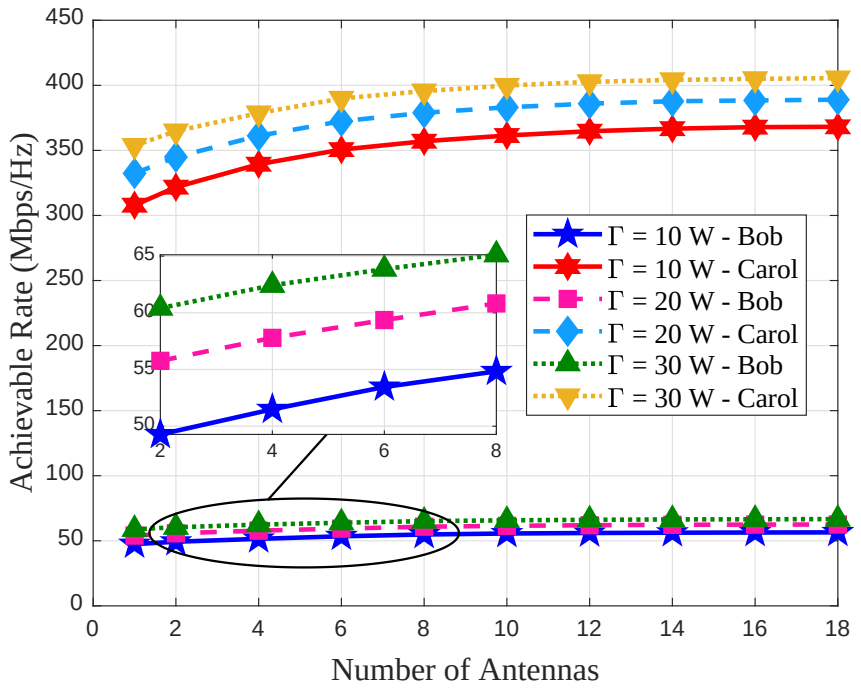

Fig. 2. The achievable rate versus the number of antennas. The study considers three cases: $\Gamma = 10$, $\Gamma = 20$, and $\Gamma = 30$.

The illustration of total AoI versus the number of antennas $M$ is shown in Fig. 3. By employing beamforming techniques, Alice can enhance channel capacity and freshly transmit packets. On the other hand, the constraints (21f) and (21g) represent the relation between the achievable data rate and the AoI for Bob and Carol for specific packet sizes. Since increasing the number of antennas increases the achievable rate, users's AoI improves, too. In addition, a higher power budget corresponds to a lower AoI, too.

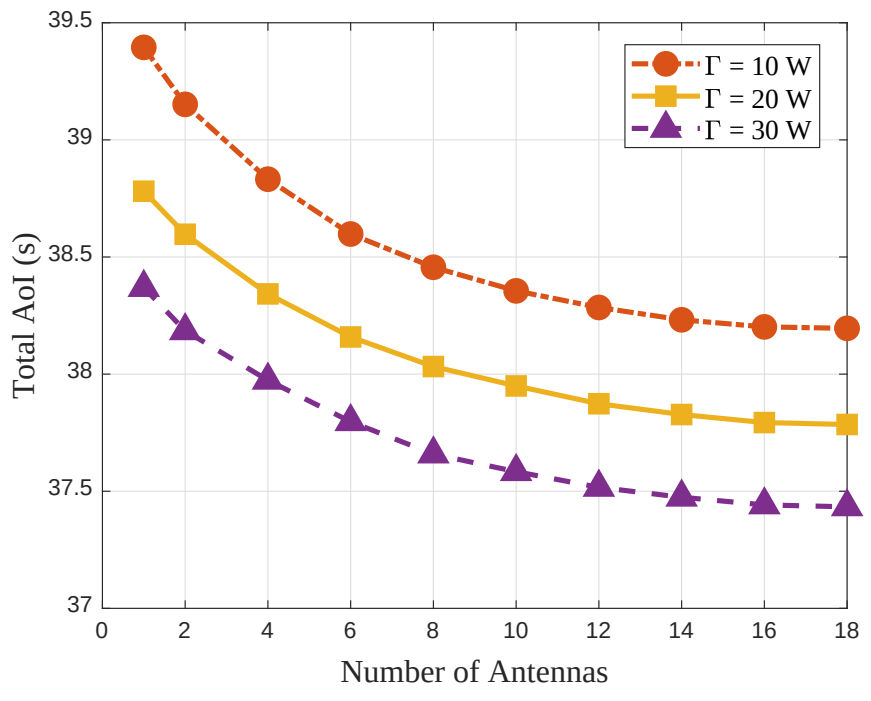

Fig. 3. The total AoI of users versus the number of antennas. The study considers three cases: $\Gamma = 10$, $\Gamma = 20$, and $\Gamma = 30$.

### B. Impact of the different covertness requirements level

Fig. 4 demonstrates the achievable covert rate versus different covertness requirements (21d). According to constraint (21d), by increasing $\epsilon$, Eve's detection error rate decreases. Therefore, Alice can allocate more power to Bob while maintaining covert transmission and improving the $R_b$. Furthermore, an increase in Alice's power budget improves Bob's performance.

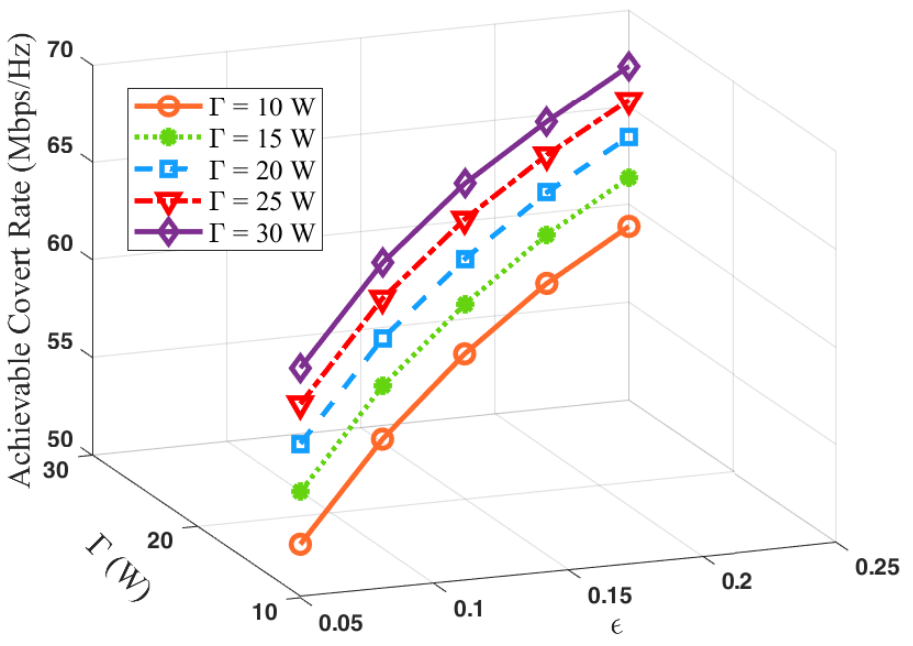

Fig. 4. The achievable covert rate versus $\epsilon$ for different transmit power of Alice $\Gamma$.

The impact of covertness requirements $\epsilon$ on the total AoI is studied in Fig. 5. As $\epsilon$ increases, the achievable covert rate increases, while the achievable public rate decreases. Based on constraints (21f) and (21g), there is an inverse relevance between the achievable rate and the AoI for specific packet sizes. Consequently, since the achievable public rate is dominant, a reduction in Carol's rate increases the total AoI. Furthermore, increasing the power budget from $\Gamma = 10$ w to $\Gamma = 30$ w enables Alice to allocate additional power resources to Carol and Bob, thereby improving the users' AoI.

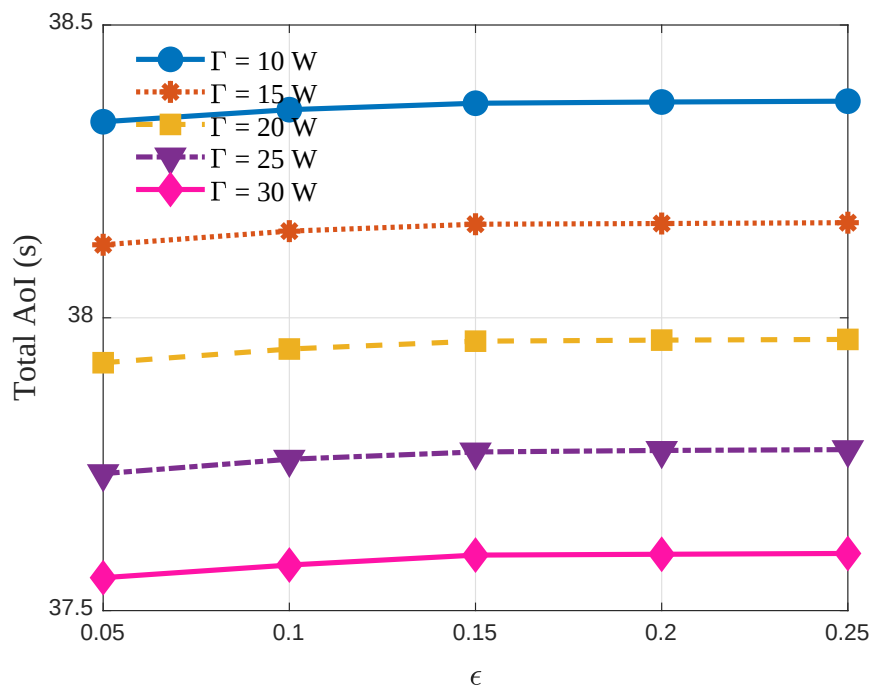

Fig. 5. The total AoI versus $\epsilon$ for different transmit power of Alice $\Gamma$.

### C. Impact of the different covert packet size

To ensure covert communication, Alice serves Carol continuously while only serving Bob upon request in each time slots that meet the constraints, such as covertness and QoS. Therefore, by increasing $S_b$ the number of allocated time slots to serve Bob increases, too. As depicted in Fig. 6, the blue part represents the number of time slots in which Alice serves only Carol, while the red part indicates the number of time slots during which Alice serves both users by adopting the PD-NOMA technique. As a result by increasing $S_b$, the average minimum detection error rate $\xi^*$ increases, too.

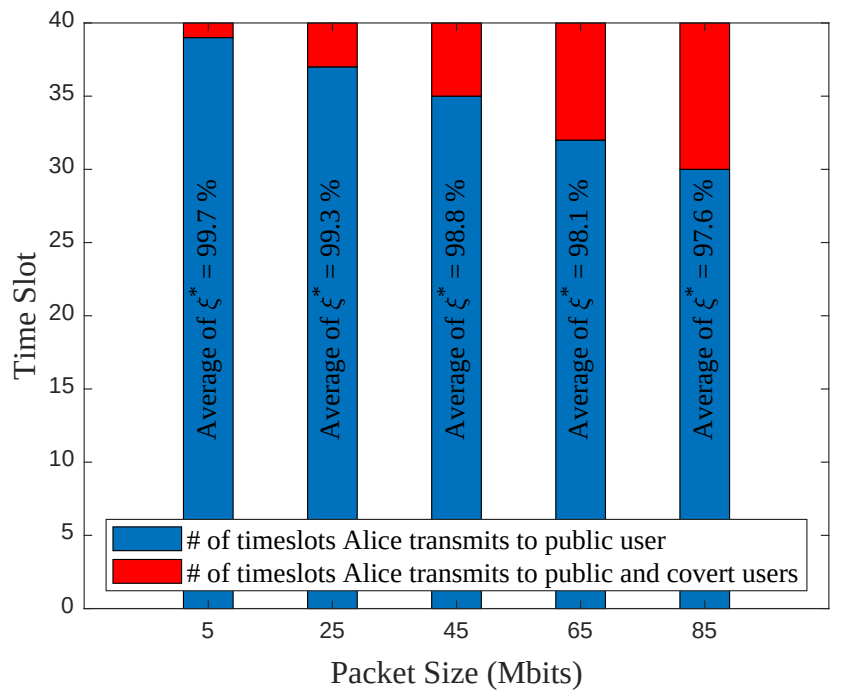

Fig. 6. An analysis of time slots for different covert packet size $S_b$.

We evaluate the covertness communication in the UAV-assisted proposed system in Fig. 7. With considering constraint (21d), Alice jointly trajectory and beamforming design to minimize the user's AoI while ensuring covert communication. In the absence of constraint (21d), Alice employs a similar strategy regardless of Eve's presence. Consequently, this may enable Eve to detect the covert transmission, resulting in the average of $\xi^*$ exceeding the guard line.

We investigate the impact of covert packet size on the achievable covert rate in Fig 8 and compare the performance of the proposed PD-NOMA system against orthogonal multiple access (OMA) as a benchmark. In the PD-NOMA framework, we consider two scenarios: one with the constraint (21d) and one without it. In the absence of

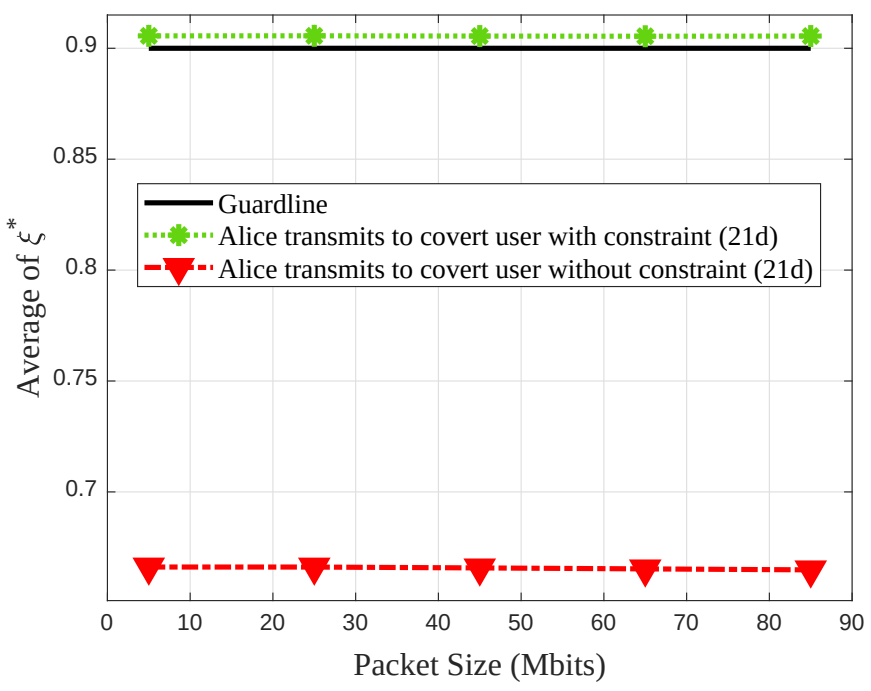

Fig. 7. The average detection error rate $\xi^*$ for different covert packet size $S_b$.

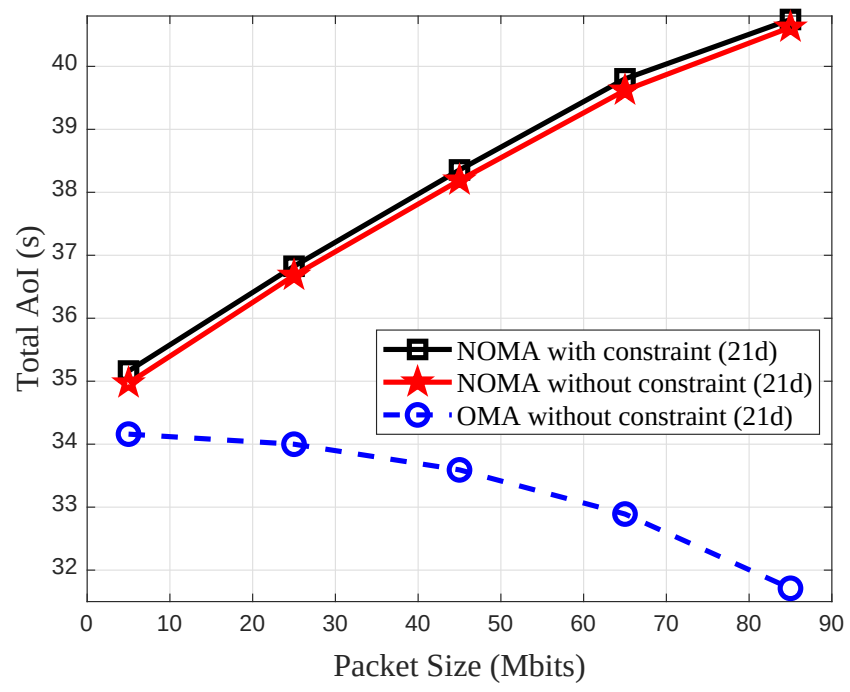

Fig. 9. Total AoI for different covert packet size $S_b$.

(21d), Alice is more flexible in her transmission strategy. Therefore, the achievable rate in this scenario is slightly higher than when applying (21d). This marginal increase indicates that our resources have not been wasted by integrating covert communication into UAV-assisted networks while the covertness is guaranteed. The OMA technique assigns each timeslot exclusively to Carol or Bob. Hence, increasing the packet size $S_b$ leads to Alice serving Bob in more time slots, and Carol's rate is not achievable. Therefore, the OMA depicts a decreasing achievable rate compared to the PD-NOMA scenarios. Consequently, the achievable rate for OMA decreases compared to the PD-NOMA scenarios [35].

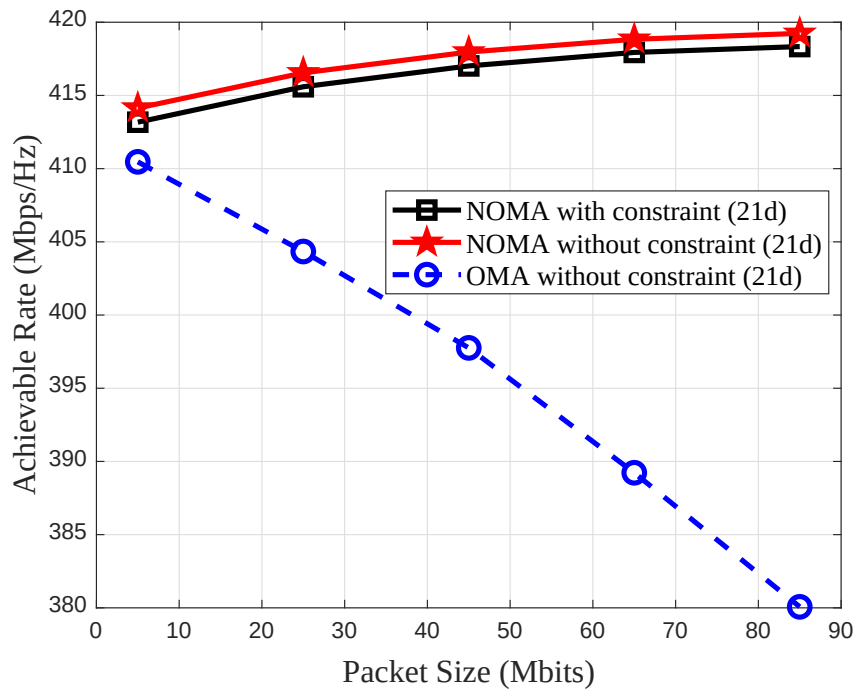

Fig. 8. The Achievable rate for different covert packet size $S_b$.

As mentioned above, increasing the covert packet size causes an increase in the number of time slots that Alice serves Bob. Consequently, the total AoI for users employing the PD-NOMA technique also increases. The PD-NOMA with constraint (21d) limits the achievable rate and leads to a higher AoI compared to the PD-NOMA technique without (21d). Conversely, when utilizing the OMA technique, whereby Alice transmits data to Carol or Bob in an orthogonal manner, the total AoI decreases in this case. The obtained results are demonstrated in Fig. 9.

### *D. Impact of the covert parameters on the trajectory design*

We present Fig. 10 to demonstrate that covertness is guaranteed from the perspective of an aerial Eve even through Alice's flying path for different covert packet sizes such as $S_b = 25$ and $S_b = 85$ Mbit. Due to the dominant time slots that Alice serves only Carol, the designed paths for the different covert packet sizes are mostly similar. However, an insignificant difference arises from the time slots in which Alice jointly serves Carol and Bob using the PD-NOMA technique. The upper and lower boxes illustrate the area and stopping points where Alice serves Bob for a given minimum required packet sizes, for example with $S_b = 85$ Mbit it takes 10-time slots, while the lower box displays for $S_b = 25$ Mbit it takes 3-time slots. The achievable covert rate for larger packets is higher.

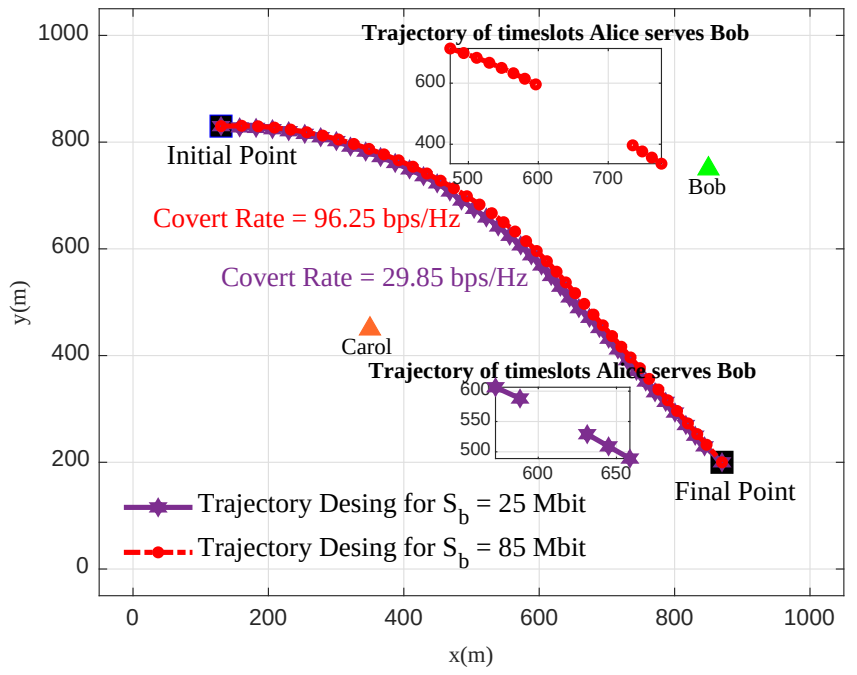

Fig. 10. Trajectory design for different covert packet size.

In this study we illustrate the achievable covert rate and the total AoI versus the covertness requirement for the following schemes: 1) Trajectory Design Path: is obtained by the proposed solution; 2) Assumption path: is obtained by the straight line connecting the initial and final points, 3) Randomly Path: is obtained by the random determination of Alice's location.

In Fig. 11, increasing $\epsilon$ results in a smaller lower bound for the covertness constraint (21d). This enables Alice to serve Bob with less limitation and increases the achievable covert rate $R_b$ for all three schemes. In the trajectory design path scheme, Alice optimally flies closer to Bob, which results in a higher $R_b$.

In Fig 12, similar to Fig 5, an increase in $\epsilon$ results in a smooth rise in the total AoI. This simulation result aligns with the theoretical analysis that reveals an inverse relationship between the achievable rate and AoI for a given packet size. Consequently, the trajectory design path scheme achieves a lower total AoI which is more desirable.

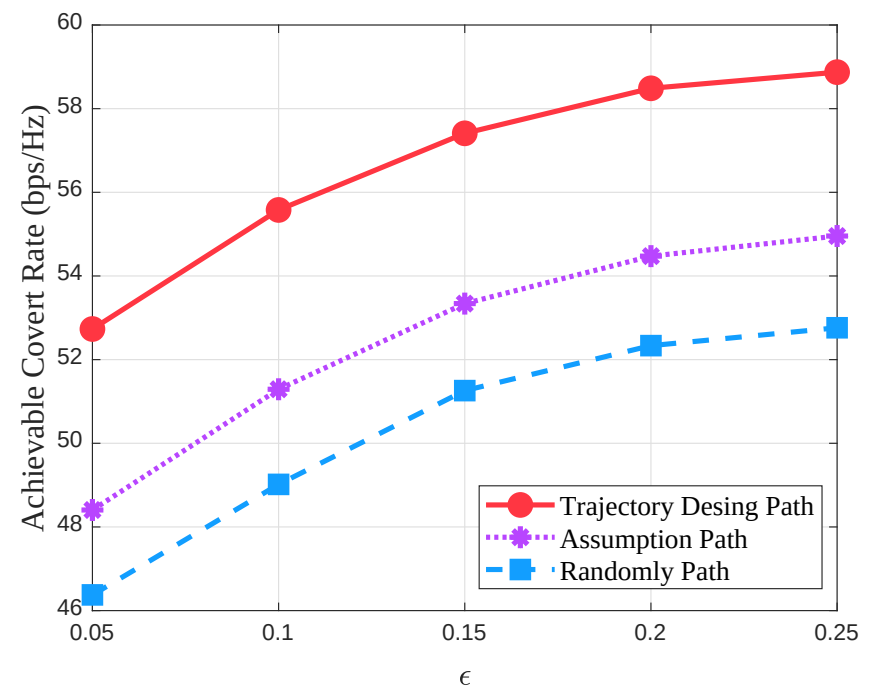

Fig. 11. Achievable covert rate versus $\epsilon$ with different flying paths.

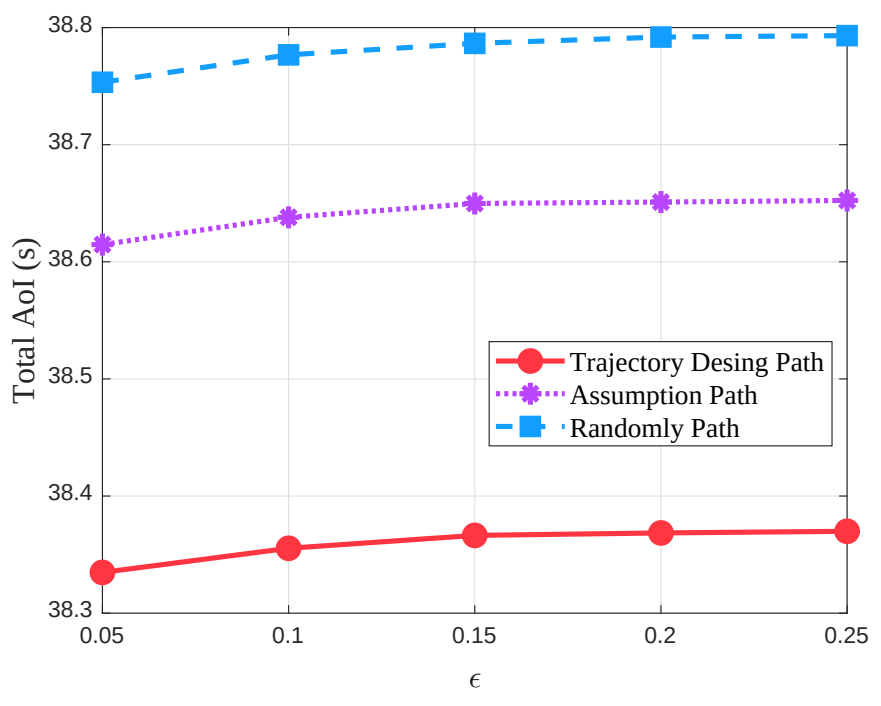

Fig. 12. Total AoI versus $\epsilon$ with different flying paths.

## V. Conclusion

This paper investigated the UAV-assisted covert communication using a beamforming technique in the PD-NOMA system. The problem of minimizing the total AoI was formulated by jointly optimizing the UAV trajectory and beamforming design, and the user's AoI. Despite considering the aerial eavesdropper as the worst case with the capability to make a decision on the detection threshold and his distance to UAV, we guaranteed the covert communication with some assumptions: 1) UAV-enabled beamforming which leads to uncertainties for detection error rate, 2) the public user covered the covert user with the PD-NOMA technique. To tackle the non-convex problem, the original problem was decoupled into three subproblems, AoI Optimization, Trajectory Design Optimization, and beamforming optimization which are solved by developing an alternating optimization solution. Numerical results demonstrated the impact of the main design parameters on the UAV-assisted covert communication system with desirable AoI. The significant performance of the PD-NOMA technique compared to OMA on the achievable rate and user's AoI was highlighted. Additionally, the assumption that "UAV continuously serves the public user while serving the covert user upon request" is well-studied.

## Appendix A
## Proof of Lemma 1

Based on the (13) and (17), the false alarm probability at Eve during the $n$-th time slot can be expressed as:

$$\mathbb{P}_{\rm FA}[n] = \int_{\tau[n]-\sigma_e^2}^{\infty} \frac{1}{\varpi_c[n]} e^{-\frac{z}{\varpi_c[n]}} dz, \ \forall n,$$
$$= \frac{1}{\varpi_c[n]} \left[ -\varpi_c[n] e^{-\frac{z}{\varpi_c[n]}} \right] \Big|_{\tau[n]-\sigma_e^2}^{\infty} = e^{\frac{\sigma_e^2 - \tau[n]}{\varpi_c[n]}}. \quad (41)$$

To derive the distribution function of the miss detection probability, we employ the moment-generating function (MGF), which represents the expected value of the exponential function of the random variable. Let $X$ be a random variable, the MGF of $X$, denoted by $\phi_X(s) = E\{e^{sX}\}$. If $s$ is a continuous random variable, the following relation between the MGF of $\phi_X(s)$ and the two-sided Laplace transform of its probability density function $f_X(x)$ holds $\phi_X(s) = L\{f_X(x)\}|_{s\to -s}$. Hence, the exponential distribution of random variable $X$ with parameter $\lambda$ is $\phi_X(s) = \frac{\lambda}{\lambda - s}$.

With assumption of $\upsilon[n] = |\alpha_c[n]|^2 + |\alpha_b[n]|^2$ and $|\alpha_c[n]|^2$ and $|\alpha_b[n]|^2$ are i.i.d, the MGF of $\upsilon[n]$ is:

$$\phi_\upsilon(s) = \phi_{|\alpha_c[n]|^2}(s)\phi_{|\alpha_b[n]|^2}(s)$$
$$= \frac{\frac{1}{\varpi_c[n]}}{\frac{1}{\varpi_c[n]} - s} \frac{\frac{1}{\varpi_b[n]}}{\frac{1}{\varpi_b[n]} - s} = \frac{\frac{1}{\varpi_c[n]}\frac{1}{\varpi_b[n]}}{\left(\frac{1}{\varpi_c[n]} - s\right)\left(\frac{1}{\varpi_b[n]} - s\right)}$$
$$= \frac{\frac{1}{\varpi_b[n]}}{\frac{1}{\varpi_b[n]} - \frac{1}{\varpi_c[n]}} \frac{\frac{1}{\varpi_c[n]}}{\frac{1}{\varpi_c[n]} - s} + \frac{\frac{1}{\varpi_c[n]}}{\frac{1}{\varpi_c[n]} - \frac{1}{\varpi_b[n]}} \frac{\frac{1}{\varpi_b[n]}}{\frac{1}{\varpi_b[n]} - s}. \quad (42)$$

Consequently:

$$f_\Upsilon(\upsilon) = \frac{\frac{1}{\varpi_c[n]}\frac{1}{\varpi_b[n]}}{\frac{1}{\varpi_b[n]} - \frac{1}{\varpi_c[n]}} \left( e^{-\frac{\upsilon[n]}{\varpi_c[n]}} - e^{-\frac{\upsilon[n]}{\varpi_b[n]}} \right)$$
$$= \frac{1}{\varpi_c[n] - \varpi_b[n]} \left( e^{-\frac{\upsilon[n]}{\varpi_c[n]}} - e^{-\frac{\upsilon[n]}{\varpi_b[n]}} \right). \quad (43)$$

According to (14) and (43), the miss detection probability at Eve for the $n$-th time slot is determined in (44). Therefore, based on (12), (41), and (44) the detection error rate $\xi[n]$ at Eve for $\tau[n] > \sigma_e^2$ region, is achieved as:

$$\xi[n] = \frac{\varpi_b[n]}{\varpi_c[n] - \varpi_b[n]} \left[ e^{\left(\frac{\sigma_e^2 - \tau[n]}{\varpi_b[n]}\right)} - e^{\left(\frac{\sigma_e^2 - \tau[n]}{\varpi_c[n]}\right)} \right] + 1. \quad (45)$$

## Appendix B
## Proof of Lemma 2

*Detection Error Rate Optimization:* As mentioned above, aerial Eve is ambitious to minimize the detection error rate $\xi^*[n]$ at each time slot. Corresponding to the achieved detection error rate at (19), $\xi[n]$ is always equal to one for $\tau[n] < \sigma_e^2$ region and this is the worst case

$$\mathbb{P}_{\rm MD}[n] = \int_0^{\tau[n]-\sigma_e^2} \frac{1}{\varpi_c[n]-\varpi_b[n]} \left(e^{-\frac{\upsilon[n]}{\varpi_c[n]}} - e^{-\frac{\upsilon[n]}{\varpi_b[n]}}\right) d\upsilon = \frac{1}{\varpi_c[n]-\varpi_b[n]} \left[-\varpi_c[n] e^{-\frac{\upsilon[n]}{\varpi_c[n]}} + \varpi_b[n] e^{-\frac{\upsilon[n]}{\varpi_b[n]}}\right] \Big|_0^{\tau[n]-\sigma_e^2},$$
$$= \frac{1}{\varpi_c[n]-\varpi_b[n]} \left[-\varpi_c[n]\left(e^{-\frac{\left(\tau[n]-\sigma_e^2\right)}{\varpi_c[n]}} - 1\right) + \varpi_b[n]\left(e^{-\frac{\left(\tau[n]-\sigma_e^2\right)}{\varpi_b[n]}} - 1\right)\right],$$
$$= \frac{\varpi_b[n] e^{\frac{\sigma_e^2-\tau[n]}{\varpi_b[n]}} - \varpi_c[n] e^{\frac{\sigma_e^2-\tau[n]}{\varpi_c[n]}}}{\varpi_c[n]-\varpi_b[n]} + 1. \quad (44)$$

$$\frac{\partial \xi[n]}{\partial \tau[n]} = \frac{B[n]}{C[n]-B[n]} \left[-\frac{d_e^2[n]}{\mu_0 B[n]} e^{\left(\frac{d_e^2[n]\left(\sigma_e^2-\tau[n]\right)}{\mu_0 B[n]}\right)} + \frac{d_e^2[n]}{\mu_0 C[n]} e^{\left(\frac{d_e^2[n]\left(\sigma_e^2-\tau[n]\right)}{\mu_0 C[n]}\right)}\right], \quad (47)$$

$$\frac{\partial \xi[n]}{\partial d_{ae}[n]} = \frac{B[n]}{C[n]-B[n]} \left[\frac{2d_{ae}[n]\left(\sigma_e^2-\tau[n]\right)}{\mu_0 B[n]} e^{\left(\frac{d_{ae}^2[n]\left(\sigma_e^2-\tau[n]\right)}{\mu_0 B[n]}\right)} - \frac{2d_{ae}[n]\left(\sigma_e^2-\tau[n]\right)}{\mu_0 C[n]} e^{\left(\frac{d_{ae}^2[n]\left(\sigma_e^2-\tau[n]\right)}{\mu_0 C[n]}\right)}\right]. \quad (48)$$

for Eve. For $\tau[n] > \sigma_e^2$ region, we note that $\xi[n]$ is a function of two variables: the detection threshold $\tau[n]$, and Eve's location $\mathbf{l}[n]$ at each time slot $n$. Eve attempts to detect the covert transmission by minimizing his distance from Alice $d_e[n]$ while determining the optimal detection threshold. Hence, we re-expressed the constraint (21d) as the following optimization problem to find the optimal error detection rate, for $\tau[n] > \sigma_e^2$ region:

$$\min_{\tau[n], d_e[n]} \xi[n] \quad (46a)$$
$$\text{s.t.} d_e[n] \geqslant d_{\min}, \quad (46b)$$

where (49b) is the collision avoidance constraint of Eve and Alice. In general, the optimal solution for the minimization problem and obtaining the corresponding minimum detection error rate $\xi^*[n]$ is partial derivatives. We take a partial derivative for $\tau[n]$ and $d_e[n]$ and set them equal to zero i.e., $\frac{\partial \xi[n]}{\partial \tau[n]} = \frac{\partial \xi[n]}{\partial d_e[n]} = 0$. Corresponding to (7), and $\varpi_k[n] = \sum_{m=1}^{M} |h_{e,m}[n]|^2 \sigma_{k,m}^2[n]$, we rewrite $\xi[n] = \frac{B[n]}{C[n]-B[n]}\left[e^{\left(\frac{d_e^2[n]\left(\sigma_e^2-\tau[n]\right)}{\mu_0 B[n]}\right)} - e^{\left(\frac{d_e^2[n]\left(\sigma_e^2-\tau[n]\right)}{\mu_0 C[n]}\right)}\right] + 1$, where $B[n] = \sum_{m=1}^{M} \sigma_{b,m}^2[n]$, and $C[n] = \sum_{m=1}^{M} \sigma_{c,m}^2[n]$. Therefore, $\frac{\partial \xi[n]}{\partial \tau[n]}$ and $\frac{\partial \xi[n]}{\partial d_e[n]}$ are calculated at (47) and (48), respectively. By setting $\frac{\partial \xi[n]}{\partial \tau[n]} = 0$, the optimal detection threshold is obtained as $\tau^*[n] = \sigma_e^2 - \ln\left(\frac{C[n]}{B[n]}\right)^{\left(\frac{B[n]C[n]\mu_0}{(B[n]-C[n])d_e^2[n]}\right)}$. In addition, by setting $\frac{\partial \xi[n]}{\partial d_e[n]} = 0$, and assuming $\tau^*[n]$, the optimal value of the error detection rate is independent of the distance between Alice and Eve $d_e[n]$. However, to detect the optimal detection error rate, an aerial Eve using a radiometer detects the maximum received signal energy $T_e[n]$. Hence, Eve is faced with the following optimization problem:

$$\max_{d_e[n]} T_e[n] \quad (49a)$$
$$\text{s.t.} d_e[n] \geqslant d_{\min}. \quad (49b)$$

We express the equivalent object of the optimization problem (49) according to the assumption of large-scale LoS path loss link's model between Alice and Eve [36], as follows:

$$\max_{d_e[n]} T_e[n] = \max_{d_e[n]} E\{|y_e[n]|^2\} \cong \max_{d_e[n]} \frac{E\{|y_e|^2\}}{E\{|y_a|^2\}}$$
$$= \max_{d_e[n]} \frac{P_e}{P_a} = \max_{d_e[n]} \left(\frac{\sqrt{G}\lambda_0}{4\pi d_e[n]}\right)^2 \quad (50a)$$

where $\cong$ is congruent symbol, and $\sqrt{G}$ is the product of the transmit and receive antenna field radiation patterns in the LOS direction with the signal wavelength $\lambda_0$. It is evident that the maximum error detection rate is attained when the distance between Alice and Eve is minimized, specifically within the allowed flying distance for the two UAVs, $d_e[n] = d_{\min}$. Consequently, the corresponding minimum detection error rate $\xi^*[n]$ associated with the optimal detection threshold $\tau^*[n] = \sigma_e^2 - \ln\left(\frac{C[n]}{B[n]}\right)^{\left(\frac{B[n]C[n]\mu_0}{(B[n]-C[n])d_e^2[n]}\right)}$, and the optimal distance $d_e^*[n] = d_{\min}$, is expressed as $\xi^*[n] = \frac{B[n]}{C[n]-B[n]}\left[\left(\frac{C[n]}{B[n]}\right)^{\frac{C[n]}{B[n]-C[n]}} - \left(\frac{C[n]}{B[n]}\right)^{\frac{B[n]}{B[n]-C[n]}}\right] + 1$.